# Relation between the $\alpha$-relaxation and the Johari-Goldstein $\beta$-relaxation of a component in binary miscible blends of glass-formers

K.L. Ngai

*Naval Research Laboratory, Washington DC 20375-5320, USA*

S. Capaccioli

*Dipartimento di Fisica and INFM (UdR Pisa), Università di Pisa, Via Buonarroti 2, I-56127, Pisa, Italy*

**Abstract**

It is well known that the $\alpha$-relaxation of each component in a miscible mixtures of two glass-formers has its own dynamics, which change with the composition of the blend. Lesser known are the corresponding change of the Johari-Goldstein (JG) $\beta$-relaxation and its relation to the $\alpha$-relaxation. Previously, in neat glass-formers, the relaxation time $\tau_{JG}$ of JG $\beta$-relaxation was identified with the independent relaxation time $\tau_0$ of the coupling model. The correspondence between $\tau_0$ and $\tau_{JG}$ was supported by analysis of experimental data of many glass-formers. In this work, this correspondence between $\tau_0$ and $\tau_{JG}$ of a component in binary mixtures and the relation between $\tau_0$ and $\tau_\alpha$ of the coupling model are used to generate predictions of the simultaneous changes of $\tau_\alpha$ and $\tau_{JG}$ of the component on varying the composition of the mixture. The predictions are in accord with the experimental data of the component 2-picoline in mixtures with either tri-styrene or *ortho*-terphenyl by T. Blochowicz and E.A. Rössler, Phys.Rev.Lett. in press (2004).

## 1. Introduction

Explaining the changes of dynamics of a component in a binary miscible mixture of two glass-formers from that in the neat state is a challenging problem. A successful explanation based on the extension of a model or theory for a neat glass-former is most desirable. There are many studies of miscible mixtures of two glass-formers on the changes of the primary $\alpha$-relaxation of either components[1,2,3,4,5]. Seldom seen is the study of the effects of mixing on both the primary $\alpha$-relaxation and the secondary $\beta$-relaxation in parallel, particularly when the secondary relaxation of one of the glass-formers is intermolecular in origin (i.e., a Johari-Goldstein relaxation)[6,7,8]. There are few examples. They include dielectric measurements of mixtures of water with glycol oligomer[9], alcohol[10], glycerol[11], sorbitol[12], and glucose[13]; mixtures of glycerol with sorbitol[14]; mixtures of 2-piocline with tri-styrene[15,16]. For polymers, there are blends of poly(vinyl methyl ether) with poly(2-chlorstyrene)[17], poly($n$-butyl methacrylate-stat-styrene) copolymers[18,19], and poly(4-vinylphenol)/poly(ethyl methacrylate) blends[20]. These studies reported change of dynamics of both the primary and the secondary relaxation of a component in binary blends. For a theoretical interpretation of the dynamics in these works, one needs an approach that can address not only the change of the $\alpha$-relaxation dynamics but also that of the secondary relaxation. Among secondary relaxations[21], the Johari-Goldstein (JG) $\beta$-relaxation[6,7,8] is the most important for glass transition because it is considered to be the precursor of the $\alpha$-relaxation. Certainly, the secondary relaxations observed in totally rigid molecular glass-formers are appropriately called JG $\beta$-relaxations after the researchers who found them. A more stringent definition of JG $\beta$-relaxation extended to non-rigid glass-formers has recently been proposed in Reference [21]. As far as we know, among theories or models of dynamics of neat glass-formers, the coupling model[22,23,24,25] is the only approach that has been extended to address the $\alpha$-relaxation and the JG $\beta$-relaxation of a neat glass-former, as well as the $\alpha$-relaxation dynamics of a component in a binary miscible blend. Naturally we apply it to construct a model of the component dynamics of a blend that address simultaneously the $\alpha$-relaxation and the J-G $\beta$-relaxation coming from a component, and their changes with blend composition. The predictions of this extended blend model are compared with the rich experimental dielectric relaxation data of 2-picoline mixed with either tri-styrene or ortho-terphenyl, published recently by Blochowicz and Rössler[15,16].

## 2. Coupling model interpretation of $\alpha$-relaxation in binary mixtures

The coupling model (CM)[22,23,24,25] emphasizes the many-body nature of the $\alpha$-relaxation dynamics of a neat glass-former $A$ through the intermolecular coupling of the relaxing species with others in its environment. The many-body dynamics are heterogeneous[26] and give rise to the Kohlrausch-Williams-Watts (KWW) stretch exponential correlation function,

$$\phi_A(t) = \exp[-(t/\tau_{A\alpha})^{1-n_A}]. \qquad (1)$$

Here $(1-n_A) \equiv \beta_A$ is the fractional KWW exponent, and $n_A$ is the coupling parameter in the CM. The larger the intermolecular coupling, the larger is $n_A$. The utility of the CM is attributed to the relation,

$$\tau_{A\alpha} = [t_c^{-n_A} \tau_{A0}]^{\frac{1}{1-n_A}}, \tag{2}$$

between the KWW relaxation time, $\tau_{A\alpha}$, to the independent relaxation time, $\tau_{A0}$, via the KWW exponent and $t_c$. The latter is the crossover time from independent relaxation to the many-body KWW relaxation and has the approximate value of $2\times10^{-12}$ s for molecular liquids[27]. The CM merely accounts for the effect of the complex many-molecule dynamics, slowing the relaxation from $\tau_{A0}$ to $\tau_{A\alpha}$ via Eq.(2). The dependences of the relaxation on temperature, entropy and volume, enter first into $\tau_{0A}$, and are amplified in $\tau_{A\alpha}$ because of the raising to the power given by the superlinear exponent, $1/(1-n_A)$, in Eq.(2). Evidences that $\tau_{0A}$ already senses the specific volume (or free volume) and/or entropy (or configurational entropy) come from that of $\tau_\beta$ of the JG relaxation. They include (1) the relaxation strength $\Delta\varepsilon_\beta$ as well as the relaxation time $\tau_\beta$[28] of the JG relaxation showing changes of temperature dependence at $T_g$; (2) the shift of $\tau_\beta$ to longer times by physical aging[29]; and (3) the large difference in $\tau_\beta$ depending on the temperature and pressure combination paths that take a glass-former from the equilibrium liquid to two different glassy states at the same final temperature and pressure[30].

Perhaps the first published model addressing the component $\alpha$-dynamics of binary mixtures A/B was by an application of the CM[31,32,33,34,35,36]. This approach extends the coupling model for neat glass-formers to mixtures by incorporating other sources of heterogeneity of dynamics in mixtures, due to the intrinsic mobility differences of the components (i.e., $\tau_{0A}$ and $\tau_{0B}$ are different) and to the local compositional heterogeneity coming from concentration fluctuations. The intrinsic mobility ($\tau_{A0}$ or $\tau_{B0}$) depends on the local composition. The $\alpha$-relaxation dynamics ($\tau_A$ or $\tau_B$) of a component ($A$ or $B$) in a mixture is determined by its chemical structure, as well as by the local environment, since the latter governs the intermolecular coupling and the many-molecule dynamics determining the $\alpha$-relaxation. The environments for either component, say $A$, are not identical due to composition fluctuations. There is a distribution of environments $\{i\}$ of the $A$ molecules, which in turn engenders a distribution of independent relaxation times $\{\tau_{A0i}\}$ and coupling parameters $\{n_{Ai}\}$ for the $A$ molecules.

To fix ideas, we consider from now on that the glass transition temperature of the neat $A$ glass-former is much lower than that of the neat $B$ glass-former, such that the intrinsic mobility of $A$ is significantly higher than $B$, or $\tau_{A0} \ll \tau_{B0}$. From the standpoint of a molecule $A$ in the mixture $A_{1-x}B_x$, the partial replacement of $A$ molecules by the less mobile $B$ molecules with different chemical structure and size in its environment $i$ will enhance the intermolecular constraints on its motion, and thereby increase its coupling parameter $n_{Ai}$ over and above $n_A$ of the neat $A$ glass-former. All $n_{Ai}$ in the distribution are larger than the coupling parameter $n_A$ of the neat $A$ glass-former, *i.e.*,

$$n_{Ai} > n_A, \tag{3}$$

The differences between $n_{iA}$ and $n_A$ become larger when there are more less-mobile $B$ molecules in the mixture. Eqs.(1) and (2) applies to each $i$ in the distribution. For each $i$ the $\alpha$-relaxation correlation function of $A$ is given by

$$\phi_{Ai}(t) = \exp[-(t/\tau_{A\alpha i})^{1-n_{Ai}}], \tag{4}$$

where $n_{Ai}$ is its coupling parameter, $\tau_{A\alpha i}$ its cooperative $\alpha$-relaxation times, and they are again related by the equation,

$$\tau_{\alpha Ai} = [t_c^{-n_{Ai}} \tau_{0Ai}]^{\frac{1}{1-n_{Ai}}}. \tag{5}$$

The observed dielectric response of all $A$ molecules in the mixture is the superposition of the one-sided Fourier transforms of Eq.(4), each weighed by the probability of the occurrence of $i$ in the distribution $\{i\}$[31,32,33,34,35,36]. Local composition richer in the less-mobile $B$ has larger $n_{Ai}$. In Eq.(5), the exponent $(1- n_{Ai})$ is the dominant quantity that determines $\tau_{A\alpha i}$. Consequently $\tau_{A\alpha i}$ of environment $i$ with larger $n_{Ai}$ is shifted to longer time and has a stronger temperature dependence. This effect causes broadening on the low frequency side of the dispersion which, in extreme cases, leads to a reversal of the asymmetry of the loss peak found in neat $A$ glass-former. Instead of skewing towards high frequency as in the one-sided Fourier transforms of Kohlrausch functions, the dispersion is altered to skewing towards lower frequency. Blends of PVME/PS and PVE and PIP exemplify this feature[31,32,33,34]. The loss peak in the dielectric and dynamic mechanical spectrum in these blends is unusually broad, and is strongly skewed toward lower frequency.

The frequency dispersion is not the emphasis of this work and hence not discussed any further. Instead we focus on the loss peak frequency, $f_{\alpha,max} \equiv 1/(2\pi\tau_{\alpha,max})$, which is determined by the contribution from the most probable ones $\hat{i}$ in the distribution of environments in the mixture $A_{1x}B_x$. Let us denote their coupling parameter by $\hat{n}_A$, the independent relaxation time by $\hat{\tau}_{A0}$, the $\alpha$-relaxation time by $\hat{\tau}_{A\alpha}$, and the correlation function by

$$\hat{\phi}_A(t) = \exp[-(t/\hat{\tau}_{A\alpha})^{1-\hat{n}_A}]. \tag{6}$$

Since $i = \hat{i}$ is just a special case, $\hat{\tau}_{A\alpha}$ can be calculated by Eq.(5) and it is given by the expression

$$\hat{\tau}_{A\alpha} = [t_c^{-\hat{n}_A}\hat{\tau}_{A0}]^{\frac{1}{1-\hat{n}_A}} \equiv \left(\frac{\hat{\tau}_{A0}}{t_c}\right)^{(\frac{\hat{n}_A}{1-\hat{n}_A})}\hat{\tau}_{A0}. \tag{7}$$

On increasing $x$, the concentration of the less-mobile component B, $\hat{n}_A$ increases due to enhanced intermolecular constraints. Since $\hat{n}_A$ appears in the exponent in Eq.(7), it is the principal cause of the increase of $\hat{\tau}_{A\alpha}$ with $x$. Because $\hat{i}$ has the highest probability of occurrence, the one-sided Fourier transform of Eq.(6) is largely responsible for determining the maximum of the observed $\alpha$-loss peak frequency of component $A$ in the mixture. Thus the experimentally determined $\tau_{A\alpha,max}$ should correspond to the calculated $\hat{\tau}_{A\alpha}$ from Eq.(7). In the limit of $x \to 0$, the mixture $A_{1-x}B_x$ becomes the neat glass-former $A$, and Eqs.(6) and (7) are reduced to Eqs.(1) and (2) respectively.

Since $t_c \approx 2$ ps is very short, in most experiments the ratio $(\hat{\tau}_{A0}/t_c)$ is much larger than unity in the entire temperature range of investigation. It follows immediately from Eq.(7) that $\hat{\tau}_{A\alpha}$ is much longer than $\hat{\tau}_{A0}$. The exponent, $[\hat{n}_A/(1-\hat{n}_A)]$, in Eq.(7) follows $\hat{n}_A$ to increase with $x$, and is principally responsible for the rapid increase of $\hat{\tau}_{A\alpha}$ with increase of $x$ at constant temperature.

### 3. Coupling model interpretation of JG $\beta$-relaxation in mixtures

A recent advance of the coupling model for a neat glass-former $A$ is a description of the evolution of dynamics with increasing time[24,25] from (*i*) the nearly constant loss in the short-time caged regime; (*ii*) the local Johari-Goldstein $\beta$-relaxation with its time $\tau_{JG}$ identified with the independent (primitive) relaxation time $\tau_{A0}$; (*iii*) the increasing probability of successful independent relaxations and concomitant continuous development of many-body cooperative dynamics until finally (*iv*) the fully cooperative $\alpha$-dynamics take hold and the KWW function becomes applicable. The rationale for identifying $\tau_{A0}$ with $\tau_{JG}$ is that they both are the relaxation times of the local and independent (non-cooperative) relaxation which is the precursor of the terminal cooperative $\alpha$-relaxation. Experimental data on various glass-forming substances[24,25,28,37,38] show remarkably good correspondence between $\tau_{A0}$ calculated by Eq.(2) and $\tau_{JG}$ from experiment,

$$\tau_{JG}(T) \approx \tau_{A0}(T) \tag{10}$$

This correspondence between the observed JG relaxation time and the independent relaxation time of component $A$ should continue to hold in the mixture $A_{1-x}B_x$, according to the coupling model. For mixtures, Eq.(10) has to be rewritten as

$$\tau_{JG}(T) \approx \hat{\tau}_{A0}(T) . \tag{11}$$

After identifying $\hat{\tau}_{A\alpha}$ with the experimentally determined $\alpha$-relaxation time $\tau_{A\alpha,\max}$ of the mixture, the prediction (11) can be tested by calculating $\hat{\tau}_{A0}$ from $\tau_{A\alpha,max}$ and $\hat{n}_A$ via Eq.(7) with the formula

$$\hat{\tau}_{A0} = (t_c)^{\hat{n}_A} (\tau_{A\alpha,\max})^{1-\hat{n}_A} . \tag{12}$$

There are many glass-formers $A$, which in the neat state do not show a resolved JG $\beta$-relaxation either in the equilibrium liquid state or the glassy state. Examples, such as glycerol[39,40], propylene glycol[41] propylene carbonate[39,42], cresol phthalate dimethyl ether[43], 2-picoline[15,16], and tri-styrene[15,16] show no secondary relaxation whatsoever in the dielectric loss spectra at any temperature. There is however the excess wing appearing on the high frequency side of the $\alpha$-loss peak. There are various evidences[15,16,44,45,46,47] that this excess wing is the JG $\beta$-relaxation hidden under the dominant $\alpha$-loss peak because $\tau_{JG}$ is not much shorter than $\tau_{A\alpha}$[21,24,25,37]. From the coupling model equations (1) and (10) and $t_c=2\times10^{-12}$ s, the separation between the two relaxations (in seconds) is given by

$$\log_{10}\tau_{A\alpha} - \log_{10}\tau_{JG} = n_A(\log\tau_{A\alpha} + 11.7) \qquad (13)$$

This relation indicates that the separation is small if the $n_A$ in the exponent $(1-n_A)$ of the KWW function used to fit the $\alpha$-relaxation of neat glass-former $A$. In fact all the glass-formers $A$ showing an excess wing but no other secondary relaxation have smaller $n_A$.

Now consider a mixture of such a glass-former $A$ with $B$ with higher $T_g$. The separation between the two relaxations in the mixture, both originating from the $A$ molecules, is given via Eqs.(12) and (13) by

$$\log_{10}\tau_{A\alpha,\max} - \log_{10}\tau_{JG} = \hat{n}_A(\log\tau_{A\alpha,\max} + 11.7). \qquad (14)$$

We have seen that $\hat{n}_A > n_A$, and $\hat{n}_A$ increases monotonically with the concentration of $B$ molecules in the mixture. It follows from Eq.(14) that in the mixtures the separation between the JG $\beta$-relaxation and $\alpha$-relaxation of the $A$ component increases monotonically with the concentration of $B$ molecules. Hence, at sufficiently high concentrations of $B$, the JG $\beta$-relaxation of $A$ will be resolved. For any mixture in which the JG $\beta$-relaxation of $A$ is resolved, we can use Eq.(14) to calculate $\hat{n}_A$ from the experimental values of $\tau_{A\alpha,\max}$ and $\tau_{JG}$. This course of action is tantamount to using one coupling model prediction on the JG $\beta$-relaxation (Eq.14) to support the description of component $\alpha$-dynamics in mixtures principally by the change of the coupling parameter of the $\alpha$-relaxation, $\hat{n}_A$, in Eq.7.

## 4. Application to 2-picoline in mixtures with tri-styrene or OTP

### (a) JG $\beta$-relaxation of 2-picoline in mixtures with tri-styrene or OTP

There are dielectric relaxation measurements on mixtures with 2-picoline with tri-styrene or 2-picoline with ortho-terphenyl (OTP)[15,16,48]. In both cases 2-picoline plays the role of glass-former $A$ that has only an excess wing but not a resolved JG $\beta$-relaxation. The dipole moments of tri-styrene and ortho-terphenyl are small compared with 2-picoline, and thus the observed dielectric spectra of the mixtures are essentially entirely coming from the relaxations of the 2-picoline molecules. At about 20% tri-styrene, the excess wing changes to exhibit a shoulder indicating the emergence of JG $\beta$-relaxation[15,16]. Above 30% tri-styrene, the JG $\beta$-relaxation are clearly seen in the loss spectra as resolved peaks. Their relaxation time $\tau_{JG}$ can be obtained directly from the isothermal dielectric loss data as the reciprocal of the angular frequency, $2\pi\nu_{JG}$, at the maximum of the JG $\beta$-loss peak. The distance between the $\alpha$-relaxation and the JG $\beta$-relaxation of 2-picoline, $(\log_{10}\tau_{A\alpha,\max} - \log_{10}\tau_{JG})$, increases with percentage of tri-styrene for any chosen fixed value of $\log_{10}\tau_{A\alpha,\max}$. For example, Fig.2 of Reference (15) show such trend for mixtures containing 60, 50, 40 and 25% 2-picoline at $\nu_{A\alpha,\max} \approx 10^{-2}$ Hz. The trend is in accord with Eq.(14) and the expected monotonic increase of $\hat{n}_A$ with the concentration of tri-styrene molecules in the mixture. One may recall that the remarkable increase of the $\alpha$-relaxation time $\hat{\tau}_{A\alpha}$ or $\tau_{A\alpha,\max}$ of 2-picoline when mixed with tri-styrene is explained in the coupling model by the monotonic increase of $\hat{n}_A$ via Eq.(7). Such proposed increase of $\hat{n}_A$ now has an independent check by calculating $\hat{n}_A$ using another prediction of the coupling model (i.e. Eq.14) and a separate experimental quantity of 2-picoline (i.e., the separation distance between the $\alpha$-relaxation and the JG $\beta$-relaxation). This check has been carried out, using the dielectric loss data shown in Fig.2 of Ref.(15) and additional data from Ref.(16). Some of the isothermal dielectric loss data of the mixtures containing 50, 40 and 25% 2-picoline in tri-styrene as well as 50 and 30% 2-picoline in ortho-terphenyl show well defined $\alpha$-loss and JG $\beta$-loss peaks and allow us to determine directly their peak frequencies $\nu_{A\alpha,\max}$ and $\nu_{JG}$ and the corresponding relaxation times $\tau_{A\alpha,\max}$ and $\tau_{JG}$, and hence $(\log_{10}\tau_{A\alpha,\max} - \log_{10}\tau_{JG})$, without using any arbitrary assumed procedure to deduce them. Next, Eq.(14) is applied to determine $\hat{n}_A$. The results of $\hat{n}_A$ are shown in Table 1 together with $n_A$=0.36 of neat 2-picoline. Approximately the same values of $\hat{n}_A$ are obtained from isothermal data available for the same mixture at different temperatures and $\tau_{A\alpha,\max}$ or $\nu_{A\alpha,\max}$, showing consistency of the results. The large and monotonic increase of the coupling parameter $\hat{n}_A$ of 2-picoline (in the most probable environments of the mixtures) with increasing content of tri-styrene is evident by inspection of the entries in Table 1 under $\{n_i\}$.

The treatment of the isothermal dielectric loss data of 5% picoline requires more explanation. At low temperatures of 170 K and below, when $\tau_{A\alpha,\max}$ becomes so long that the entire $\alpha$-relaxation of picoline has moved out of the experimental frequency window, the JG $\beta$-relaxation was observed as a broad symmetric loss peak. The Arrhenius temperature dependence of $\tau_{JG}$ is given by $3\times10^{-15}\exp[27.27$ (kJ/mol)$/RT]^{15,16}$. However, at the higher temperatures of 223 K and above, the $\alpha$-loss peaks were seen but the measurements had not been carried out to sufficiently high frequencies to find the corresponding JG $\beta$-relaxation. In order to obtain the quantity, $(\log_{10}\tau_{A\alpha,\max} - \log_{10}\tau_{JG})$, at the temperature $T_{Ag}$ defined by $\log_{10}\tau_{A\alpha,\max}(T_{Ag})=10^2$ s, we extrapolate the Arrhenius dependence of $\tau_{JG}$ at temperatures below $T_{Ag}$ to $T_{Ag}$ to determine $\log_{10}\tau_{JG}(T_{Ag})$. Approximately we have, at $T_{Ag}\approx 220.3$ K, $\log_{10}\tau_{JG}(T_{Ag})=-8.06$. From the latter and Eq.(14), we deduce $\hat{n}_A=0.735$. The coupling parameter of the 5% picoline is large and is the largest among all the mixtures studied by Blochowicz et al. (see Table 1). This is certainly expected because of the preponderance of the less-mobile tri-styrene molecules seen by the picoline in this mixture.

The data of blends with more than 50% picoline do not show a resolved JG loss peak[15,16] and there is large uncertainty in the determination of $\tau_{JG}$ using any arbitrary fitting procedure. For this reason, they are not considered. It is worthwhile to remind the reader that, unlike neat glass-formers, $\hat{n}_A$ can be determined directly from the dispersion of the $\alpha$-loss peak in mixtures because of the broadening caused by concentration fluctuations. This is unfortunate because they are the key parameters in the application of the coupling model to component $\alpha$-relaxation dynamics in binary mixtures, and one would like to obtain their values and see how they change with mixing. Nevertheless, by using another prediction of the coupling model on the JG $\beta$-relaxation, we are able to calculate $\hat{n}_A$ and the favorable results obtained support the premise of the coupling model for the component $\alpha$-relaxation dynamics in binary mixtures. After $\hat{n}_A$ is known, one can proceed to fit the $\alpha$-relaxation dispersion in the mixture by summing over the contributions from all environments and assuming a plausible Gaussian distribution of $\{n_i\}$. In the process, one can determine the width of the distribution of $\{n_i\}$. The task is not carried out here because this is outside the focus of the present work.

**(b) The ratio $E_\beta/(RT_{Ag})$ from the $\alpha$- and JG $\beta$-relaxation of 2-picoline in mixtures**

At temperatures below $T_g$, the most probable relaxation times of all secondary relaxations have Arrhenius temperature dependence[15,16]. In particular, for the JG $\beta$-relaxation, we have

$$\tau_{JG}(T) = \tau_\infty \exp(E_\beta/RT), \quad T \leq T_g, \tag{15}$$

where $\tau_\infty$ is the prefactor, $E_\beta$ the activation enthalpy, and $R$ the gas constant. A remarkable empirical relation of the JG $\beta$-relaxation to the glass transition temperature $T_g$ of neat glass-formers given by

$$E_\beta \approx 24RT_g, \tag{16}$$

was found by Kudlik et al.[49,50,51]. Although the relation is only approximate and there are deviations, it is a remarkable finding which suggest some connection exists between the JG $\beta$-relaxation and the $\alpha$-relaxation. A relation between $E_\beta$ and $T_g$ does exist as a consequence of Eq.(13) when considered at $T=T_g$. It is given by

$$E_\beta/(RT_g) = 2.303[(1-n_A)\log_{10}\tau_{A\alpha}(T_g) - 11.7n_A - \log_{10}\tau_\infty]. \tag{17}$$

Most measurements of secondary relaxations are obtained by means of dielectric relaxation spectroscopy, where $T_g$ is conveniently defined as the temperature at which the dielectric relaxation time $\tau_\alpha$ reaches an arbitrarily long time, typically $10^2$ s. Following this convention, on substituting $\tau_\alpha(T_g)=10^2$ s and $t_c=2$ ps into Eq.(17), we arrive at the expression,

$$E_\beta/(RT_g) = 2.303(2 - 13.7n_A - \log_{10}\tau_\infty). \tag{18}$$

We have previously shown that the expression on the right-hand-side of Eq.(18) when evaluated gives a good account of the experimental value $E_\beta/(RT_g)$ for many neat glass-formers, not only those that obey the approximate empirical rule of Eq.(16) but also others that deviate significantly from it[52]. In other words, Eq.(18) is a more general relation than Eq.(16). The ratio, $E_\beta/(RT_g)$, is larger for a smaller $n_A$ and a shorter $\tau_\infty$. It turns out that for some neat glass-formers, the ones that have a larger $n_A$ or small KWW exponent, $(1-n_A) \equiv \beta_A$, also have a shorter $\tau_\infty$ for their JG $\beta$-relaxation. The compensating effects of $n_A$ and $\tau_\infty$ make possible the ratio, $E_\beta/(RT_g)$, to be approximately constant for a number of neat glass-formers.

We now test the quantitative prediction of $E_\beta/RT_g$ given by Eq.(18) against the data of the $\alpha$- and the JG $\beta$-relaxation in the mixtures of picoline with tri-styrene and with ortho-terphenyl[15,16]. The analogue of Eq.(18) for picoline in the blends is

$$E_\beta/(RT_{Ag}) = 2.303(2 - 13.7\hat{n}_A - \log_{10}\tau_\infty), \quad (19)$$

where now $T_{Ag}$ is defined by $\hat{\tau}_{A\alpha}(T_{Ag}) = 10^2$ s. The experimental values of the prefactor $\tau_\infty$ are taken from Refs.(15) and (16) and $\hat{n}_A$ has previously been determined and given in Table 1. The theoretical values of $E_\beta/(RT_{Ag})$ calculated via Eq.(18) are in good agreement with the experimental values for all blends (see Table 1), except for the blend of 50% picoline with tri-styrene which shows some slight deviation. The discrepancy in this case is likely due to uncertainties incurred in determining $\tau_{JG}$ of the 50% picoline mixture. The uncertainties arise because $v_{JG}$, the frequency of the maximum of the dielectric loss of the JG relaxation, at most temperatures in the glassy state were not determined directly but were deduced by some fitting procedure[15,16].

The ratio $E_\beta/(RT_{Ag})$ for the 25, 40 and 50% picoline mixtures are not far from the value of 25, found for many neat glass-formers[49,50,51,52]. One may notice from Table 1 for these three mixtures that the decrease of $\hat{n}_A$ is accompanied by the increase of $\log_{10}\tau_\infty$. The near constancy of $E_\beta/(RT_{Ag})$ arises from compensating changes of $\hat{n}_A$ and $\tau_\infty$ in Eq.(19). The 5% picoline mixture is an exception having a much smaller $E_\beta/(RT_{Ag})$ equal to 14.5. However, the calculated value is still in good agreement with it.

### (c) Non-Arrhenius temperature dependence of $\tau_{JG}$

Like in neat glass-formers, the most probable $\alpha$-relaxation time of the component $A$ in a binary mixture, $\tau_{A\alpha,max}$, has non-Arrhenius temperature dependence at temperature sufficiently high such that the $\alpha$-dynamics of the component $A$ is taking place in equilibrium. The non-Arrhenius temperature dependence is often approximately described by the Vogel-Fulcher-Tammann-Hesse (VFTH) equation,

$$\tau_{A\alpha,\max}(T) = C\exp[D/(T-T_0)], \qquad (20)$$

where $C$, $D$ and $T_0$ are constants. On cooling the mixture, $\tau_{A\alpha,max}$ becomes long enough that the substructure of the component $A$ in the blend falls out of equilibrium at some temperature, $T_{Ag}$. At temperatures below $T_{Ag}$ the substructure is frozen and the temperature dependence of $\tau_{A\alpha,max}$ is given by the Arrhenius equation,

$$\tau_{A\alpha,\max}(T) = \tau_{A\alpha\infty}\exp(E_{A\alpha}/RT), \qquad (21)$$

where $\tau_{A\alpha\infty}$ is the prefactor, $E_{A\alpha}$ the activation enthalpy, and $R$ the gas constant. The coupling model relations, Eqs.(11) and (12), lead immediately to the prediction of the corresponding temperature dependence of $\tau_{JG}$:

$$\tau_{JG} \approx \hat{\tau}_{A0} = (t_c)^{\hat{n}_A}(\tau_{A\alpha,\max})^{1-\hat{n}_A} = (t_c)^{\hat{n}_A}(C)^{1-\hat{n}_A}\exp[\frac{(1-\hat{n}_A)D}{T-T_0}], \qquad T > T_{Ag} \quad (22)$$

and

$$\tau_{JG} \approx \hat{\tau}_{A0} = (t_c)^{\hat{n}_A}(\tau_{A\alpha,\max})^{1-\hat{n}_A} = (t_c)^{\hat{n}_A}(\tau_{A\alpha\infty})^{1-\hat{n}_A}\exp[\frac{(1-\hat{n}_A)E_{A\alpha}}{RT}], \quad T < T_{Ag} \quad (23)$$

From these it is clear that the Arrhenius temperature dependence of $\tau_{JG}$ of component $A$ at temperatures below $T_{Ag}$ will give way to the VFTH dependence above $T_{Ag}$. The same prediction was made on neat glass-formers, which are Eqs.(22) and (23) after replacing $\hat{n}_A$ therein by $n_A$. However, $n_A$ of most neat small molecular glass-formers are less than 0.5. The distance between the $\alpha$- and the JG relaxation frequencies are not large enough to permit an unequivocal determination of $\tau_{JG}$ at temperatures above $T_g$. A case in point is neat sorbitol. In spite of it having $n_A$=0.52, some arbitrary fitting procedure[53,54] has to be used to determine $\tau_{JG}$ at temperatures above $T_g$. Only by subjecting sorbitol to high pressure[Error! Bookmark not defined.] was it possible to determine $\tau_{JG}$ directly and showed that the temperature dependence of $\tau_{JG}$ is in accord with Eqs.(20) and (21). In this respect, the study of the JG relaxation in mixtures, like picoline/tri-styrene, has an advantage. We have seen that $\hat{n}_A$ of picoline can be increased significantly by increasing the content of the less mobile tri-styrene. The mixtures with 25% and 40% picoline have respectable large values of $\hat{n}_A$ equal to 0.62 and 0.48 respectively (see Table 1). The JG loss peak is clearly observed at temperatures above and below $T_{Ag}$, from which $\tau_{JG}$ is *directly* determined. $\tau_{JG}$ indeed shows[15,16] the change from the Arrhenius dependence of Eq.(23) to a stronger dependence at some temperature near $T_{Ag}$, consistent with Eq.(22). Hence the dielectric data of the picoline component in the 25% and 40% picoline mixtures offer direct evidence that the Arrhenius temperature dependence of $\tau_{JG}$ in the glassy state does not continue to hold in the equilibrium liquid state. In mixtures with more than 40% picoline, the $\tau_{JG}$'s were obtained by a fitting procedure[15,16] that involves some assumption. Nevertheless, the deduced $\tau_{JG}$'s also show the change of temperature dependence across $T_{Ag}$'s of the blends.

## 5. Conclusion

The coupling model has been applied to describe the $\alpha$-relaxation dynamics of a component in binary mixtures of two glass-formers. This extension of the coupling model is one among several proposed models for the $\alpha$-relaxation dynamics of a component in mixtures. There are already some experimental data of the $\alpha$-relaxation of a component that distinguish the coupling model from other models[31,32,33,34,55]. In this work, we have extended the coupling model further to give a parallel description of the $\alpha$-relaxation and the JG $\beta$-relaxation of a component in the mixture. Such extension has several quantitative predictions on the JG $\beta$-relaxation time and its temperature dependence, and the relation of these quantities to the $\alpha$-relaxation of the same component. These predictions are consistent with the $\alpha$-relaxation and the JG $\beta$-relaxation of 2-picoline in binary mixtures with either tri-styrene or ortho-terphenyl.

**Acknowledgment**

The research was supported by at NRL by the Office of Naval Research, and at the Università di Pisa by I.N.F.M.. We thank Prof. E. Rössler for sending a copy of Reference 16.

**Table 1**. Experimental and calculated parameters of the $\alpha$-relaxation and the JG $\beta$-relaxation of 2-picoline in binary mixtures with either tri-styrene or ortho-terphenyl. There is a difference between the calorimetric $T_g$ given in Refs.(15) and (16) from the dielectric $T_{Ag}$ obtained from the dielectric relaxation data by the definition $\tau_{A\alpha,\max}(T_{Ag})=10^2$ s. Whenever the difference is small, we enter $(E_\beta/RT_{Ag})_{\exp}$ given in Refs (15) and (16), which were computed by using the calorimetric $T_g$. In the case of the 50% 2-picoline mixture, the calorimetric $T_g$ differs by about 5 degrees. When we compute $(E_\beta/RT_{Ag})_{\exp}$ by using the dielectric $T_{Ag}$, the result indicted by an asterisk is a bit smaller.

| % of 2-picoline in tri-styrene | $\log_{10}(\nu_{A\alpha,\max}/\text{Hz})$ | $\log_{10}(\nu_{JG}/\text{Hz})$ | $\hat{n}_A$ from Eq.(14) | $\tau_\infty$ | $(E_\beta/RT_g)_{\exp}$ | $(E_\beta/RT_g)_{\text{cal}}$ from Eq.(17) |
|---|---|---|---|---|---|---|
| 5 | -2.8 | 7.26 | 0.735 | $3\times10^{-15}$ | 14.5 | 14.9 |
| 25 | -2.4 | 5.8 | 0.62 | $3\times10^{-17}$ | 23.1 | 23.1 |
| 25 | -3.4 | 5.5 | 0.62 | $3\times10^{-17}$ | 23.1 | 23.1 |
| 25 | -4.3 | 5.2 | 0.62 | $3\times10^{-17}$ | 23.1 | 23.1 |
| 40 | -2.0 | 4.1 | 0.47 | $5\times10^{-16}$ | 26.0 | 25.0 |
| 40 | -3.8 | 3.2 | 0.48 | $5\times10^{-16}$ | 26.0 | 24.7 |
| 40 | -0.7 | 4.4 | 0.44 | $5\times10^{-16}$ | 26.0 | 26.0 |
| 50 | -3.9 | 2.3 | 0.41 | $10^{-13}$ | 25.9 24.9* | 21.6 |
| 100 | -0.8 | N/A | 0.36 | N/A | N/A | N/A |
| % of 2-picoline in OTP | $\log_{10}(\nu_{A\alpha,\max}/\text{Hz})$ | $\log_{10}(\nu_{JG}/\text{Hz})$ | $\hat{n}_A$ from Eq.(14) | $\tau_\infty$ | $(E_\beta/RT_g)_{\exp}$ | $(E_\beta/RT_g)_{\text{cal}}$ from Eq.(17) |
| 30 | -2.4 | 4.5 | 0.52 | $6\times10^{-17}$ | 25.5 | 25.6 |
| 30 | -1.0 | 4.9 | 0.50 | $6\times10^{-17}$ | 25.5 | 26.2 |
| 50 | -3.0 | 2.5 | 0.40 | $4\times10^{-15}$ | 26.5<br>25.7[*] | 25.1<br>25.1 |
| 100 | -0.8 | N/A | 0.36 | N/A | N/A | N/A |

**References**

[1] R.E Wetton,.W.J. MacKnight, J.R. Fried, and F.E. Karasz Macromolecules, **11,** 158 (1978).

[2] A. Zetsche, F. Kremer, W. Jung, and H. Schulze, Polymer, **31,** 1883 (1990).

[3] C.M. Roland and K.L. Ngai, Macromolecules, **25,** 363 (1992).

[4] P.G. Santangelo, C.M. Roland, K.L. Ngai, A.K. Rizos, and H.J. Katerinopoulos, Non-Cryst. Solid., **172-174**, 1084 (1994).

[5] C. Svanberg, R. Bergman, P. Jacobson, and L. Börjesson, Phys.Rev.B, **66**, 054304 (2002).

[6] G.P. Johari, and M. Goldstein, J. Chem. Phys., **53,** 2372 (1970).

[7] G.P. Johari, G. Power, and J.K. Vij, J.Chem.Phys., **117**, 1714 (2002)**.**

[8] G.P. Johari. J. Non-Cryst. Solids, **307-310**, 317 (2002).

[9] N. Shinyashiki, S. Sudo, W. Abe, and S. Yagihara, J. Chem. Phys., **109**, 9843 (1998).

[10] T. Sato, A. Chiba, and R. Nozaki, J. Mol. Liq., **101**, 99 (2002).

[11] S. Sudo, M. Shimomura, N. Shinyashiki, and S. Yagihara, J. Non-Cryst. Solids, **307-310**, 356 (2002).

[12] R. Nozaki, H. Zenitani, A. Minoguchi, and K. Kitai, J. Non-Cryst. Solids, **307-310**, 349 (2002).

[13] M. Tyagi and S.S.N. Murthy, J.Phys.Chem.B to be published (2004).

[14] K. Duvvuri and R. Richert, J.Phys.Chem.B to be published (2004).

[15] T. Blochowicz and E.A. Rössler, Phys.Rev.Lett. in press (2004).

[16] T. Blochowicz, *Broadband Dielectric Spectroscopy in Neat and Binary Molecular Glass Formers*, ISBN 3-8325-0320-X, Logos Verlag, Berlin (2003). The data we utilized to bring out the relation between the $\tau_\alpha$ and $\tau_{JG}$ of 2-picoline in blends with tri-styrene, and their simultaneous changes on varying the composition of the mixture, are taken mainly from this publication.

[17] O. Urakawa, Y. Fuse, H. Hori, Q. Tran-Cong, and K. Adachi, Polymer, **42**, 765 (2001).

[18] S. Kahle, J. Korus, E. Hempel, R. Unger, S. Höring, K. Schröter, and E. Donth, Macromolecules, **30**, 7214 (1997).

[19] K.L. Ngai, Macromolecules, **32**, 7140 (1999).

[20] Shihai Zhang, Xing Jin, Paul C. Painter, and James Runt, Macromolecules, **35**, 3636 (2002).

[21] K.L. Ngai and M. Paluch, J.Chem.Phys.**120**, 875 (2004).

[22] K.L. Ngai, Comments Solid State Phys., **9**, 141 (1979).

[23] K.L. Ngai and K.Y. Tsang, Phys.Rev.E, **60**, 4511 (1999).

[24] K.L. Ngai. J. Phys.: Condens. Matter, **15**, S1107 (2003).

[25] K.L. Ngai.and M. Paluch, J. Phys. Chem. B, **107**, 6865 (2003).

[26] K.L. Ngai and R.W. Rendell, "The Symmetric and Fully Distributed Solution to a Generalized Dining Philosophers Problem", *Relaxation in Complex Systems and Related Topics*, edited by I. A. Campbell and C. Giovannella, NATO ASI Series, Vol. **222**, Plenum, New York (1990) pp. 309-316.

[27] K.L. Ngai and R.W. Rendell, in *Supercooled Liquids, Advances and Novel Applications*, J.T. Fourkas, D. Kivelson, U. Mohanty, and K. Nelson, eds., ACS Symposium Series 1997, Vol. **676**, Am.Chem.Soc., Washington, DC., Chapter 4, p.45.

[28] M. Paluch, C.M. Roland, S. Pawlus, J. Ziolo, and K.L. Ngai, Phys. Rev. Lett., **91**, 115701 (2003).

[29] D. Prevosto, S. Capaccioli, M. Lucchesi, P.A. Rolla, and K.L. Ngai, J.Chem.Phys. **120**, 4808 (2004).

[30] M. Paluch, S. Pawlus, S.Hensel-Bielowka, M. Sekula, T. Psurek, S.J. Rzoska, and S J. Ziolo, Phys. Rev. Lett., submitted (2004).

[31] C.M. Roland and K.L. Ngai, Macromolecules, **24**, 2261 (1991).

[32] C.M. Roland and K.L. Ngai, J. Rheology, **36**, 1691 (1992).

[33] C.M. Roland and K.L. Ngai, Macromolecules, **25**, 363 (1992); **33**, 3184 (2000).

[34] A. Alegria, J. Colmenero, K.L. Ngai, and C.M. Roland, Macromolecules, **27**, 4486 (1994).

[35] C.M. Roland and K.L. Ngai, Macromolecules, **28**, 4033 (1995).

[36] C.M. Roland and K.L. Ngai, J.M. O'Reilly, and J.S. Sedita, Macromolecules, **25**, 3906 (1992).

[37] K.L. Ngai, J.Chem.Phys, **109**, 6982 (1998).

[38] S. Hensel-Bielowka, M. Paluch, J. Ziolo, and C.M. Roland, J. Phys. Chem. B, **106**, 12459 (2002).

[39] P. Lunkenheimer, U. Schneider, R. Brand, and A. Loidl, Contemp. Phys. **41**, 15 (2000).

[40] K. Duvvuri and R. Richert, J.Chem.Phys. **118**, 1356 (2003).

[41] C. León, K.L. Ngai, and C.M. Roland, J. Chem. Phys. **110**, 11585 (1999).

[42] K. L. Ngai, P. Lunkenheimer, C. León, U. Schneider, R. Brand, A. Loidl, J. Chem. Phys. **115**, 1405 (2001).

[43] M. Paluch, K. L. Ngai, S. Hensel-Bielowka, J.Chem.Phys. **114**, 10872 (2001).

[44] S.H. Chung, G. P. Johari and K. Pathmanathan, J. Polym.Sci.: Part B: Polymer Phys. 24, 2655 (1986).

[45] U. Schneider, R. Brand, P. Lunkenheimer, and A. Loidl, Phys. Rev. Lett. **84**, 5560 (2000).

[46] C. Svanberg, R. Bergman, and P. Jacobson, Europhys.Lett. **64**, 358 (2003).

[47] R. Casalini and C.M. Roland, Phys.Rev.Lett. **91**, 015702 (2003).

[48] T. Blochowicz, C. Tschirwitz, S. Benkhof, and E. Rössler, J. Chem. Phys. **118**, 7544 (2003).

[49] A. Kudlik, C. Tschirwitz, S. Benkhof, T. Blochowicz and E. Rössler, Europhys.Lett. **40**, 649 (1997).

[50] A. Kudlik, C. Tschirwitz, T. Blochowicz, S. Benkhof and E. Rössler, J.Non-Cryst.Solids. **235-237**, 406 (1998).

[51] A. Kudlik, S. Benkhof, T. Blochowicz C. Tschirwitz, and E. Rössler, J.Mol.Structure **479**, 210 (1999).

[52] K.L. Ngai and S. Capaccioli, Phys.Rev.E, **69**, 0315xx (2004).

[53] T. Fujima, H. Frusawa, and K. Ito, Phys. Rev. E **66**, 031503 (2002).

[54] R. Nozaki, D. Suzuki, S. Ozawa, Y. Shiozaki, J.Non-Cryst.Solids **235-237**, 393 (2002).

[55] K.L. Ngai and C.M. Roland, Macromolecules, in press (2004).